\documentclass[a4paper,twoside]{article}
\usepackage{times}

\usepackage{soul}
\usepackage{url}
\usepackage[hidelinks]{hyperref}
\usepackage[utf8]{inputenc}
\usepackage[small,skip=4pt,belowskip=0pt]{caption}
\usepackage{graphicx}
\usepackage{calc}
\usepackage{amssymb}
\usepackage{amstext}
\usepackage{amsmath}
\usepackage{amsthm}
\usepackage{multicol}
\usepackage{apalike}
\usepackage[bottom]{footmisc}
\usepackage{SCITEPRESS}
\usepackage{booktabs}
\usepackage[compact]{titlesec}

\usepackage{balance} 

\usepackage{comment, todonotes}
\usepackage[inline]{enumitem}

\usepackage{listings}
\usepackage{algcompatible, algorithm, algpseudocode}
\algnewcommand{\LineComment}[1]{\State \(\triangleright\) #1} 
\newcommand*\Let[2]{\State #1 $\gets$ #2} 
\MakeRobust{\Call} 

\newcommand{\BibTeX}{\rm B\kern-.05em{\sc i\kern-.025em b}\kern-.08em\TeX}

\newcommand{\pre}[1]{\ensuremath{{#1}_{pre}}}
\newcommand{\goal}[1]{\ensuremath{{#1}_{goal}}}
\newcommand{\deadline}[1]{\ensuremath{{#1}_{deadline}}}
\newcommand{\priority}[1]{\ensuremath{{#1}_{prio}}}
\newcommand{\start}[1]{\ensuremath{{#1}_{start}}}
\newcommand{\duration}[1]{\ensuremath{{#1}_{duration}}}
\newcommand{\post}[1]{\ensuremath{{#1}_{post}}}
\newcommand{\context}[1]{\ensuremath{{#1}_{cont}}}
\newcommand{\effect}[1]{\ensuremath{{#1}_{eff}}}
\newfloat{execution}{htbp}{loa}
\floatname{execution}{Execution}

\usepackage{color}

\newcommand{\robol}[2]{#1}
\newcommand{\roveri}[2]{#1}

\usepackage{xpatch}
\xpatchcmd\algorithmic
  {\labelwidth 1.2em}
  {\labelwidth .7em}
  {}{\fail}

\begin{document}

\title{Real-Time BDI Agents: a model and its implementation}

\author{
  \authorname{
    Andrea Traldi,
    Francesco Bruschetti,
    Marco Robol,
    Davide Calvaresi,
    Marco Roveri,
    Paolo Giorgini
  }
  \affiliation{
    University of Trento, Trento, Italy
  }
  \email{
    \{andrea.traldi, francesco.bruschetti\}@studenti.unitn.it,
    \\
    \{marco.robol, marco.roveri, paolo.giorgini\}@unitn.it
  }
  \affiliation{
    University of Applied Sciences Western Switzerland, Sierre, Switzerland
  }
  \email{
    \{davide.calvaresi\}@hevs.ch
  }
}



\abstract{The BDI model proved to be effective for developing applications requiring high-levels of autonomy and to deal with the complexity and unpredictability of real-world scenarios. The model, however, has significant limitations in reacting and handling contingencies within the given real-time constraints. Without an explicit representation of time, existing real-time BDI implementations overlook the temporal implications during the agent's decision process that may result in delays or unresponsiveness of the system when it gets overloaded. In this paper, we redefine the BDI agent control loop inspired by well established algorithms for real-time systems to ensure a proper reaction of agents and their effective application in typical real-time domains. Our model proposes an effective real-time management of goals, plans, and actions with respect to time constraints and resources availability. We propose an implementation of the model for a resource-collection video-game and we validate the approach against a set of significant scenarios.}

\onecolumn \maketitle \normalsize  \setcounter{footnote}{0} \vfill

\sloppypar


\section{\uppercase{Introduction}}
\label{sec:introduction}
Today's applications require more and more systems capable of taking decisions autonomously in dynamic, complex and unpredictable environments. In addition, decisions have to be taken in timely fashion so to avoid that when the system starts to execute, the plan or the action is no longer necessary or appropriate in that specific situation. In other words, decisions have to be taken in real-time and this is particularly true for critical systems where a delay in a decision can compromise the life of humans. While run-time performances mainly depends on hardware, response time are obtained by adequate real-time software architectures, that are often used in critical applications to control physical systems, such as in the context of aviation or automotive systems.

To guarantee deadlines, real-time architectures estimate a computation cost and consequently allocate an appropriate computational capacity. This approach is not adopted in the state-of-art multi-agent architectures, including BDI, where real-time principles are not used and there is no way for an agent to reason about how to meet its deadlines. Moreover, agents’ architectures are thought to operate in extremely dynamic contexts in which it is often impossible to forecast all possible events and then making hard to estimate the actual computation cost. For example, consider a video-game in which agent-driven Non-Player Characters (NPCs) play with real players who can have completely unexpected behaviours. NPCs have to find solutions and take decisions with a reaction time ideally as quick as humans.

Although, BDI architectures can take decisions autonomously in complex situations, they don’t perform well in real-time interactions with humans and therefore they have strong limitations in many critical real-world scenarios. The main motivation of this is that, although frameworks such as Jade~\cite{bellifemine1999jade}, Jack~\cite{busetta1999jack}, and others~\cite{pokahr2005jadex,huber1999jam} allow for a temporal scheduling of agent's tasks (e.g., asking to perform an action periodically or at a given time), they do not use any explicit representation of time in the decision-making processes. This causes agents to overlook the temporal implications, causing delays when an overload occurs. In~\cite{DBLP:conf/paams/AlzettaGMC20}, the authors proposed a BDI-based architecture that overcomes such limitations by integrating real-time mechanisms into the reasoning cycle of a BDI agent. However, while this can guarantee the respect of time constraints at the task level (i.e., every task completes its execution within its relative deadline), the goals of the agents are not bounded to a deadline, and so the agent cannot take decisions based on temporal restrictions of goals.
AgentSpeak(RT)~\cite{DBLP:conf/atal/VikhorevAL11} allows programmers to specify deadlines and priorities for tasks. \cite{DBLP:journals/aamas/CalvaresiCMDNS21} proposes a formal model for RT-MAS. Both works do not address the scheduling of tasks/actions on a real-time environment.

In this paper, \roveri{we make the following contributions. First,}{} we propose a real-time (RT) BDI framework inspired by traditional and well established algorithms for real-time systems to ensure \robol{predictability of execution.}{a proper reaction of agents and their effective application in typical real-time domains.} \robol{We consider as primitive citizens fully integrated in a traditional BDI model concepts like computational capacity, deadline, scheduling constraints, durative actions, and periodic tasks.}{Concepts like computational capacity, deadline, scheduling constraints, durative actions and periodic tasks are fully integrated in the traditional BDI model to guarantee real-time planning and real-time tasks execution of an agents.}
\robol{This allows an agent to be aware of time constraints while deliberating and not only during the execution.
As far as our knowledge is concerned, this is the first framework that encompass all these concepts.
Second, we demonstrate the practical applicability of the proposed framework through an implementation, instantiation and validation within a resource-collection video game based on Unity, built on top of an already exiting simulator (Kronosim). This synthetic scenario is paradigmatic of several real-life applications. This\roveri{ as far as we know,}{} is the first practical implementation showing that all these concepts could be deployed in practical scenarios.}{
To demonstrate the validity of the approach, we implemented the approach and we instantiated it within a resource-collection video game implemented leveraging Unity.
We used the developed video game to validate the implemented real time BDI agents.}

The paper is structured as follows. Section~\ref{sec:baseline-game} presents the baseline. Section~\ref{sec:real-time_bdi_architecture} presents our three layers Real-time BDI architecture. Section~\ref{sec:kronosim-based_simulator} describes the simulator we build to run real-time BDI agents. \roveri{Section~\ref{sec:related_works} position this work w.r.t. the state-of-the-art.}{} Finally, in Section~\ref{sec:validation}, we discuss some scenarios we used for the validation and then we conclude the paper in Section~\ref{sec:conclusion}.


\section{\uppercase{Baseline}}
\label{sec:baseline-game}
\roveri{In this work we build on
\begin{enumerate*}[label=\roman*)]
\item the Kronosim real-time BDI simulation environment;
\item a temporal planning infrastructure;
\item and the Unity multi-platform game engine.
\end{enumerate*}
In the rest of this section we summarize the main concepts of these components.}{}

\textbf{Kronosim}\footnote{Kronosim source code is available at \url{https://github.com/RTI-BDI/Kronosim}, documentation can be found at \url{http://www.bruschettifrancesco.altervista.org/kronosim}.} is a C++ based simulation tool that combines \textit{hard real-time} concepts (e.g. responsiveness within deadline and resource computation constraints) with a classic BDI agent model to enable simulating and testing dynamic scenarios involving Real-Time BDI (RT-BDI) agents. An RT-BDI agent is able to make autonomous decisions, even in dynamic environments, and ensure time compliance. In Kronosim the RT-BDI agents have a \textit{maximum computational power} $U$~\cite{DBLP:conf/paams/AlzettaGMC20}, and leverage Earliest Deadline First (EDF) and Constant Bandwidth Server (CBS)~\cite{PAAM10} mechanisms to ensure time compliance during the execution process. The resulting algorithm allows to schedule and fairly execute, a set of \textit{"aperiodic"} and \textit{"periodic in an interval"} tasks (the former are are executed only once, the latter are repeated multiple times at predefined intervals of time). Kronosim allows for the simulation of a single RT-BDI agent, where the intentions are stored in a knowledge base that associates each intention with the set of desires it satisfies (thus it does not provide for the deliberation of new intentions to handle not yet specified desires).

\textbf{Temporal planning with PDDL~2.1}~\cite{DBLP:journals/jair/FoxL03} is a framework for
\begin{enumerate*}[label=\roman*)]
\item modeling the behavior of agents considering that actions might not be instantaneous and last some known amount of time,
\item dynamically generate time-triggered plans (i.e., sequences of actions where each action is associated with the time instant at which the action shall be scheduled, and the respective duration) for achieving a goal. \roveri{Time-triggered plans allows to represent multiple actions active at the same time, thus capturing the case where two different actions are executed in parallel by two different agents or the case where a single agent executes two actions in parallel by leveraging two different actuators in parallel to perform a task. These are our minimal requirements to represent the possible activities of the agents.}{}
\end{enumerate*}
%
Among \roveri{the possibly many}{possible} temporal planners \roveri{supporting our minimal requirements}{}, we consider OPTIC~\cite{DBLP:conf/aips/BentonCC12} which supports the generation of \roveri{time-triggered}{} temporal plans minimizing a \roveri{given}{} cost function\roveri{}{over preferences}.
\roveri{We remark that, the planning community also considered richer formalisms like e.g. PDDL 3.1~\cite{pddl31} that complements PDDL 2.1 with constraints, preferences, and other features. These features will be nice to have in the framework we will describe later on, but, as far as our knowledge is concerned, there are no tools that support all the features of PDDL 3.1. For the time being, we considered only the minimal requirements as provided by PDDL 2.1.}{}

\textbf{Unity} (\url{https://unity.com/}) is a recent multi-platform game engine, to develop graphical interface for users to interact with the game. Unity, besides aesthetics and user interface management for a game, can provide the sprite rendering, the asset store, time management, and the usage of frames as time units to drive simulations.




\section{\uppercase{The Real-Time BDI Architecture}}
\label{sec:real-time_bdi_architecture}
One of the most important feature of BDI-based agents is the ability to decide which goals to pursue and how to achieve such goals.
Traditionally, the goal deliberation problem is addressed at the agent programming level i) leveraging pre-programmed plans handled and coded by the agent developer in an error-prone ad-hoc manner; ii) without considering deployment constraints (e.g., real-time constraints). Moreover, although existing BDI solutions allow for temporal scheduling of agent's tasks, they do not use any explicit representation of time in the decision making and deliberation processes. This may cause the agent to overlook the temporal implications, and may result in unacceptable delays.

\begin{figure*}
  \centering
  \includegraphics[width=0.95\textwidth]{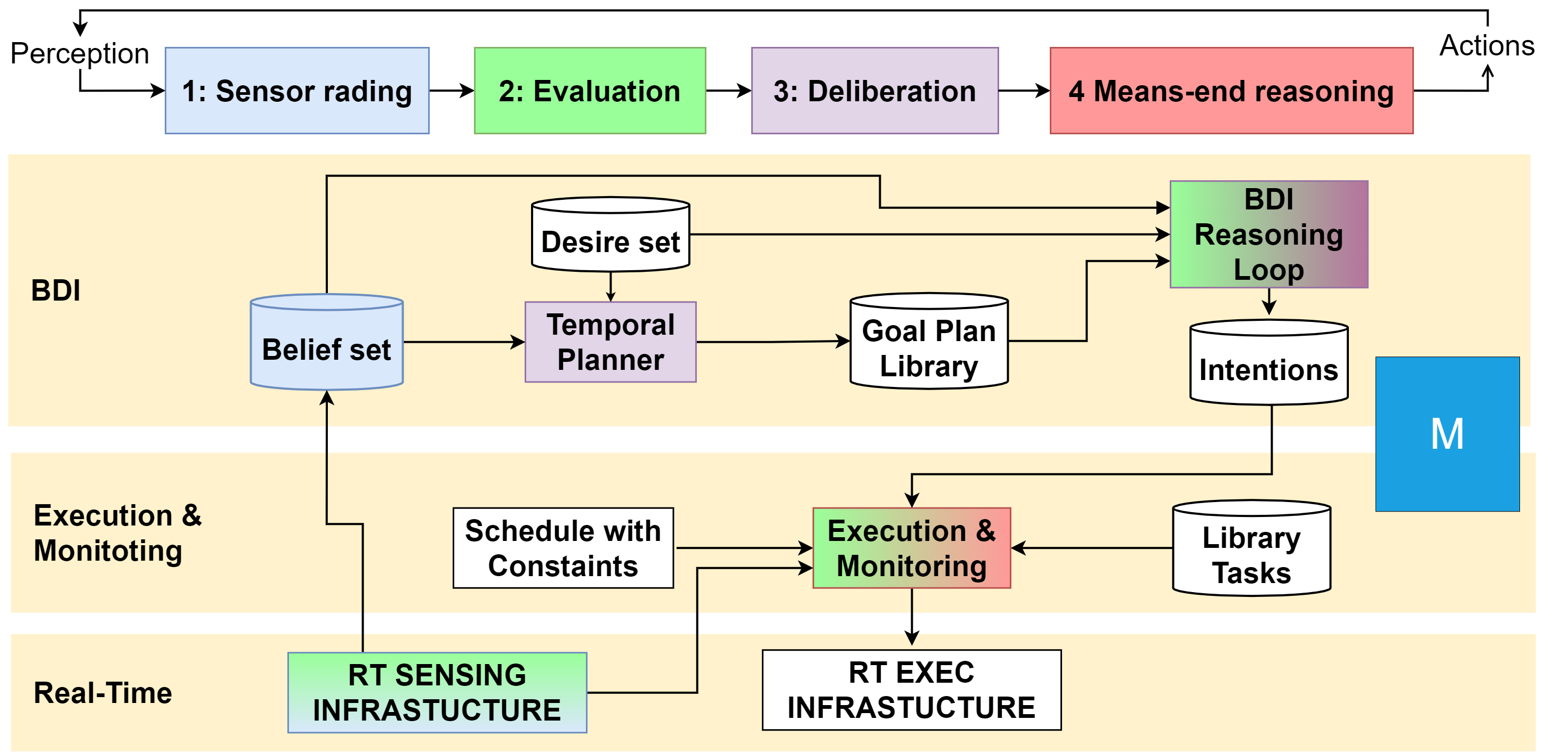}
  \caption{RT-BDI agent's architecture.}
    \label{fig:reasoning_cycle}
\end{figure*}

To avoid these limitations, we designed a three layer RT-BDI architecture, namely  \emph{BDI},  \emph{Execution and Monitoring}, and \emph{Real-Time} layers. This framework is inspired by traditional and well established algorithms to ensure prompt reaction, and encompass a reasoning cycle that consider as primitive citizens concepts like computational capacity, deadlines, scheduling constraints, periodic tasks and deliberation.
Figure \ref{fig:reasoning_cycle} shows pictorially our three layers RT-BDI architecture.

\textbf{The BDI layer} is responsible of all the high-level BDI reasoning
capabilities, i.e. handling belief, desires (goals) and intention
(temporal plans) thus implementing and extending a classical BDI
reasoning cycle.
This layer uses a model $M$ (shared also with the lower layer) to
enable for the logical reasoning (i.e. the language to represent and
reason about beliefs and environment), and to define the dictionary of
the possible actions that the agent can perform to operate on the
environment (comprehensive of the preconditions, duration, effects).
Each desire $d$ has
\begin{enumerate*}[label=\roman*)]
\item a \emph{pre-condition} \pre{d} i.e. a formula that represents
  the condition that triggers the activation of the desire itself;
\item a \emph{goal formula} \goal{d} that represents the condition to
  be achieved;
\item a \emph{deadline} \deadline{d} that represents the (relative)
  time instant we expect the goal to be achieved;
\item a \emph{priority} \priority{d} an integer representing the level
  of priority for the goal.
\end{enumerate*}
Each intention (plan) consists of either an \emph{atomic plan} (that
corresponds to a durative action to be executed), a \emph{sequential
  plan} (that enforces a sequential execution of the sub-plans, and
execution of sub-plan at $i$-th position requires successful execution
of all the $j \le i$ sub-plans), or \emph{parallel plan} (that allows
for a parallel execution of the sub-plans, with a synchronization
consisting of all the parallel sub-plans to be successfully
terminated). Each atomic plan $\pi^a$ is associated with
\begin{enumerate*}[label=\roman*)]
\item a pre-condition $\pre{\pi^a}$ (i.e. a formula that is expected
  to hold in the current state for the atomic plan to be applicable);
\item a start time $\start{\pi^a}$ (i.e. a (relative) time instant at
  which the atomic plan is expected to be started);
\item a duration time $\duration{\pi^a}$ (i.e. the expected duration
  of the atomic plan);
\item a context condition $\context{\pi^a}$ (i.e. a formula that is
  expected to hold for the entire duration of the atomic plan);
\item an effect condition $\effect{\pi^a}$ (i.e. a formula that
  specifies how the atomic plan is expected to modify the belief
  states);
\item a post-condition $\post{\pi^a}$ (i.e. a formula that specifies
  what is expected to hold when the atomic plan is terminated, and
  that entails the effect condition).
\end{enumerate*}
The BDI layer stores the desires in the \emph{Desire set}, and uses
the \emph{Goal Plan Library} to maintain an association between the
desires and a set of plans each achieving the associated
goal. Moreover, it maintains the \emph{Belief set} that represents the
current belief state of the agent (it can be seen as a map that
assigns a value to all the symbols specified in the language of the
model $M$).

The \emph{BDI Reasoning Loop} iterates over the Desire set and selects
among the desires whose preconditions hold in the current Belief set
those with highest priority and removes them from the Desire set
adding them in the set of \emph{active desires} $G$.
Then for each desire $d \in G$, it checks whether there exists a plan
achieving it in the Goal Plan Library. If no plan exists, then a
\emph{Temporal Planner} is invoked to generate one (assuming the goal
is achievable). The plan generated by the Temporal Planner is then
stored in the Goal Plan Library. At this point, there is a plan
achieving desire $d$ in the Goal Plan Library, and it is selected and
inserted in the \emph{active intentions} set $I$ for being executed by the lower layer.

\textbf{The Execution and Monitoring layer} is responsible of the execution and
monitoring of the plans to achieve the goals. In this layer, if the plan
preconditions hold the execution of the composing actions is started. The verification of the preconditions
consists in evaluating the formula expressing them w.r.t. the sensed
state. The verification of the preconditions is complemented with the
verification of real-time schedulability constraints to ensure proper
execution of the plan in a real-time environment, given the current
active plans. If such constraints are not met, then the BDI layer is
notified to activate respective further reasoning. If these
preliminary checks pass, then the execution and monitoring of the
actions in the plan starts. The monitoring verifies that the expected
effects of the different actions or the possible context conditions (that
shall hold for the whole action duration) are satisfied. If not
satisfied, the execution of the plan is aborted, and a notification is
sent to the above layer triggering further reasoning\robol{, which may include the need for re-planning and re-simulation of tasks schedulability}{}. This layer
leverage on a \emph{Library of Tasks} that maps each atomic plan into
\emph{low level periodic tasks} to be executed in the underneath RT
operating system.

\textbf{The Real-Time layer} is responsible of the execution of the
different low level tasks in a real-time operating system environment
according to classical task execution strategies of a real-time
operating systems (e.g. RT-Linux, VxWorks) 
following the scheduling policy established in the above layer.

Algorithm~\ref{alg:reasoning_cycle} reports the pseudo-code for our
revised reasoning cycle. Initially, the set of active intentions ($I$)
and active goals ($G$) are empty (lines \ref{a1l1}-\ref{a1l2}).
\begin{algorithm*}
  \caption{RT-BDI Reasoning cycle}
  \label{alg:reasoning_cycle}
  \scalebox{1}{
  \begin{minipage}{\textwidth}
  \begin{algorithmic}[1]
    \LineComment{$D$: Desire set, $P$: Goal Plan Library, $M$: Model}
    \Function{ReasoningCycle}{$D$, $P$, $M$}
    \Let{$G$}{$\emptyset$} \Comment{Active goals} \label{a1l1}
    \Let{$I$}{$\emptyset$} \Comment{Active intentions}\label{a1l2}
    \Let{$B$}{\Call{ReadSensingData}{$M$}} \Comment{Read sensing data} \label{a1l3}
    \While{(True)}\label{a1l4}
        \Let{$G$}{\Call{UpdateActiveGoals}{$G$,$B$,$D$}}\label{a1l5} \Comment{Updates $G$ with all $d^i \in D$ such that $\pre{d^i}$ hold in $B$}
        \Let{$I$,$P$}{\Call{SelectIntentions}{$B$, $G$, $I$, $P$, $D$}}\label{a1l6} \Comment{Selects intentions for each active goal}
        \Let{$I$,$G$}{\Call{RT-ProgressAndMonitorIntentions}{$B$, $I$, $G$}}\label{a1l7}\Comment{Progresses the intentions and goals}
        \Let{$B$}{\Call{ReadSensingData}{$M$}} \Comment{Read sensing data}\label{a1l8}
      \EndWhile\label{a1l9}
    \EndFunction
  \end{algorithmic}
  \end{minipage}
 }
\end{algorithm*}

The belief set $B$ representing the current knowledge of the agent
(all the symbols in the model $M$ have a value) is initialized through
reading the sensors (line \ref{a1l3}) calling \Call{ReadSensingData}{$M$}.
Then the loop starts (lines \ref{a1l4}-\ref{a1l9}), and at the beginning (line \ref{a1l5}) the set of active goals is updated through function \Call{UpdateActiveGoals}{$G$,$B$,$D$} that iterates over the desire set $D$ and adds in $G$ those desires $d \in D$ such that $\pre{d}$ holds in $B$.
Upon possible update of $G$, function \Call{SelectIntentions}{$B$,
  $G$, $I$, $P$, $D$} is invoked (line \ref{a1l6}) to update the
active intentions $I$. This function for each newly added desire $g$
first checks whether there exists in the Goal Plan Library a plan
$\pi$ such that
\begin{enumerate*}[label=\roman*)]
\item achieves $g$;
\item $\pre{\pi}$ holds in the current belief state $B$;
\item $\deadline{\pi} \le \deadline{g}$.
\end{enumerate*}
If such a plan $\pi$ exists, then it is added to $I$. If more than one plan $\pi$ exists, then the "best"~\footnote{The choice of the plan to add considers several factors, like e.g., duration, computational cost w.r.t. the other plans already in $I$.} one is selected and added to $I$.
Otherwise, a temporal planner is invoked to compute such a plan for the given desire $g$ starting from $B$ (the returned plan by construction will be such that it achieves the goal $g$, the preconditions will be satisfied in $B$, and the deadline will also be satisfied). In this case the newly generated plan will be added to $I$, and to the Goal Plan Library to enlarge the knowledge-base.
Then the set of active intentions is given to the function
{\Call{RT-ProgressAndMonitorIntentions}{$B$, $I$, $G$}} that is responsible
of executing/progressing the plans in $I$ and to update the set of
active goals $G$. In particular this functions performs the following
steps.
First, for each of the plans $\pi$ in the set of active intention $I$
it keeps track of where the respective execution is (i.e. like a
program counter in a classical program). Each plan is executed
according to a topological sorting visit of the graph representing the
plan $\pi$. If an atomic action $\pi^a$ needs to be executed, then
first it is checked if its preconditions hold in the current belief
state $B$. If it is not the case, the execution of the plan is aborted
and the information is propagated to handle the
contingency. Otherwise, the respective low-level task is activated and
inserted in the set of \emph{low level active tasks} (i.e. those low
level periodic tasks to be executed in the RT environment).
The set of low level active tasks is sorted in a \emph{low level
  active tasks list} according to a given task scheduling policy
(e.g. EDF that exhibits good behavior~\cite{calvaresi2018timing}).
The scheduling policy will consider not only the deadlines of the different low level active tasks, but also their priorities. Then the respective execution slot for each element in the low level active task list is executed in the low level RT environment according to the computed order.
At each low-level execution cycle, it is verified whether the context conditions of all the executing tasks holds, and possible anomalies are reported to higher layers to handle them.
When a low-level task terminates, it is removed from the low level active task
list, and it is checked whether the atomic task post conditions hold in the current belief state (obtained through internal calls to \Call{ReadSensingData}{$M$}), and if it is the case the frontier of the execution of the corresponding plan is updated to progress to the remaining parts of the plan. Otherwise, the task is aborted, and the problem is reported to higher levels.
Finally, at line \ref{a1l8}, the current belief state is updated, and the loop repeated.



\section{\uppercase{The Real-Time BDI Agents Simulator}}
\label{sec:kronosim-based_simulator}

In order to validate the proposed RT-BDI architecture, we
\begin{enumerate*}[label=\roman*)]
\item implemented it as an extension of the Kronosim multi-agent real-time simulator;
\item deployed and configured the simulator within a video-game scenario built on top of the Unity framework.
\end{enumerate*}
%
We extended Kronosim along three main directions.
First, we added the feature to invoke a PDDL 2.1 based deliberation engine
(we have chosen OPTIC~\cite{DBLP:conf/aips/BentonCC12})
planner to synthesize new plans. 
This functionality is seen as an external service with a corresponding API to be invoked by Kronosim to compute
a new plan.
This solution enables for plugging different deliberation engines and/or distributing the computation load in the cloud or among different computation resources if needed. 
Second, we added a proper API to communicate with a physical/dygital environment to acquire sensing data, to send actuation commands, and to monitor the respective progress.
%
%
Third, we added the possibility to add dynamically new desires to the Desire set for being considered in the BDI reasoning cycle.
Finally, we extended the framework to handle periodic tasks needed to simulate the execution of the actions within the real-time system infrastructure with a strict collaboration with the physical and/or digital system through the defined API.

In order to invoke the external planner the internal representation of the model $M$ is converted in PDDL by leveraging on the language to define the objects and predicates, and on the library of tasks to create the (durative) actions and the respective preconditions, effects, context condition and duration.
The current Belief set $B$ together with a goal $g \in G$ is converted in PDDL as well to constitute the problem file to be given to the planner for generating the plan.
The temporal plan computed by OPTIC specifies for each action the time instant the action should start, and the respective duration.
This structure is then converted in the internal representation within Kronosim as follows.
We create a parallel plan where each branch corresponds to an action in the temporal plan computed by the planner.
The time when an action shall start its execution, is used to specify a delay timing for each branch of the constructed parallel plan.
The preconditions associated to each parallel branch action are taken directly from the respective action's preconditions.
The resulting parallel plan structure preserves the semantics of the temporal plan generated by OPTIC, and simplifies its execution.
More sophisticated encoding structures could be considered for instance leveraging Behavior Trees and considering causal dependencies among the actions (but this is left for future work).

The environment and Kronosim executions operate in parallel following a kind of step-wise alternated execution: Kronosim receives updates from the environment about new desires, modification of the environment (e.g. new obstacles appears, battery level), completion of an action. Then Kronosim uses this information to monitor the execution of the active intentions $I$ and to progress them. Indeed, in Kronosim, actions execution and effects are (resp.) activated, perceived and verified in the external environment after waiting for the time of execution. If it is not possible to validate expected effects, as in the case of external interference, the goal is re-planned in the next reasoning cycle.

The communication of the components is performed using a client-server protocol based on TCP/IP, thus allowing for a separation of concerns, and for the possible distribution of the computation load among different machines.

\paragraph{Limitations.}
Our implementation suffers of the following limitations.
The Real-Time Layer only considers task scheduling and goal deadlines. It disregard plan search time and time for executing the BDI loop. This limitation could affect performances in real-world agents. Future work include a study about how to weaken this limitation by introducing more requirements on the simulation environment, as well as adopting a satisfycing temporal planner that sacrifices optimality in spite of efficiency.
However, we remark that we generate plans that already consider deadlines as first-citizens thus more suitable to be scheduled in a real-time environment.
We also remark that, the research is temporal planning is not yet mature as the one on classical planning due to the higher problem complexity.
The current implementation supports goals (and sub-goals) with associated deadlines within (manually specified) plans, and relies on the deadline information associated to compute the real-time schedule. When we execute a plan that contains goals to schedule we use the deadlines associated to the subgoals. In particular, we check whether there exists a plan that achieves the goal and meets the goal deadline. If such a plan exists we execute it. Otherwise we invoke the planner to search for a new plan respecting the associated deadline and achieving the goal. If the planner succeeds we execute the new generated plan. Otherwise we give up execution by aborting the plan.
Another limitation of the current implementation consists in the fact that the agent reacts to external events after all currently scheduled tasks have completed execution.
This is due to the underlying adopted simulator, which was not originally intended for run-time simulation, but only for simulating predefined scenarios.
%
%
Despite of these limitations, we strongly believe that our theoretical architecture, is suitable for real-time systems.




\section{\uppercase{Validating the RT-BDI Agent Architecture}}
\label{sec:validation}

\robol{
To validate our proposed solution, we \roveri{deployed it in}{consider} a video-game setting where each agent in the game executes on top of a \roveri{unique}{} real-time execution environment with limited computation capabilities, that is possibly running other real-time tasks.
This example/experiment 
is paradigmatic of real scenarios in which the agents (who in our game are assigned with goals by the players) execute their tasks on top of a \roveri{unique}{} real-time execution environment to guarantee a deterministic behavior and the required predictability of execution.
}
{To validate our proposed solution we deployed it in a video-game setting. To this extent,
}

\roveri{The implemented resource collection video game}{We implemented}\roveri{,}{a resource collection video game} called \textbf{Kronity}\footnote{Kronity source code will be released to the community in open-source after acceptance.}
, \roveri{is such that}{where}
\begin{enumerate*}[label=\alph*)]
\item the concept of \roveri{NPC}{non-player character} has the same purpose as the BDI agent model;
\item the player interaction with the game provides a simulation of an unpredictable environment\roveri{.}{;}
\end{enumerate*}
The game consists of a set of robots moving in a 2-D grid with possible obstacles that can move resources among different locations. Resources can be produced and/or consumed. The robots while moving consume fuel, and can be refueled in particular locations.

Based on this video-game we performed a detailed validation leveraging on the following paradigmatic scenarios:
\begin{enumerate*}[label=\roman*)]
\item the ability to promptly react to the happening of external events that may require re-planning some or all the current active intentions;
\item the ability to coordinate multiple agents in an unified environment;
\item the efficiency of the RT-BDI reasoning cycle;
\item the ability to learn new plans to then reuse in other situations if suitable for fulfilling new desires;
\item the handling of goal deadlines and real-time constraints.
\end{enumerate*}
In the remaining part of the section we will discuss the third and last scenarios, and we refer to the additional material for the others.\footnote{Additional documentation and a video of the simulation are available upon request.}


\begin{execution*}
\caption{Simulator Execution log.}
\label{alg:execution_logs}
\scalebox{0.8}{
\begin{minipage}{1.25\textwidth}
\begin{algorithmic}[1]
    \State{\textbf{[0]} \textsc{ReasoningCycle}: execution 1 }

    \State{\textbf{[0]} \textsc{UpdateActiveGoals}: pursue goal $G1$: "resources $R1$ and $R2$ delivered to the warehouse $W$" }

    \State{\textbf{[0]} \textsc{SelectIntentions}: available plan $P1$ ([0] $C1$ move\_up; [10] $C1$ move\_right; [20] $C1$ move\_right, ...) selected to pursue $G1$}

    \State{\textbf{[0]} \textsc{RT-ProgressAndMonitorIntentions}: $I1$($P1$: $C1$ move\_up) }

    \State{\textbf{[10]} \textsc{ReadSensingData}: $C1$ moved up }

    \State{\textbf{[10]} \textsc{ReasoningCycle}: execution 2 }

    \State{\textbf{[10]} \textsc{UpdateActiveGoals}: goal $G1$ still valid }

    \State{\textbf{[10]} \textsc{SelectIntentions}: plan $P1$ still valid, intention $I1$ still active }

    \State{\textbf{[10]} \textsc{RT-ProgressAndMonitorIntentions}: $I1$($P1$: $C1$ move\_right) }

    \State{\textbf{[15]} \textsc{Player's interaction}: A new robot "$C2$" is added to the scene }

    \State{\textbf{[20]} \textsc{ReadSensingData}: robot $C1$ has moved right and robot "$C2$" has been added to the scene }

    \State{\textbf{[20]} \textsc{ReasoningCycle}: execution 3 }

    \State{\textbf{[20]} \textsc{UpdateActiveGoals}: goal $G1$ still valid }

    \State{\textbf{[20]} \textsc{SelectIntentions}: new plan $P2$ generated, $I2$ activated based $P2$ }

    \State{\textbf{[20]} \textsc{RT-ProgressAndMonitorIntentions}: $I2$($P2$: $C1$ move\_up \& $C2$ move\_up)}

    \LineComment{... simulation progresses ...}

    \State{\textbf{[60]} \textsc{ReadSensingData}: robot "$C1$" is on "$R1$", robot "$C2$" is on "$R2$" }

    \State{\textbf{[70]} \textsc{ReasoningCycle}: execution 6 }

    \State{\textbf{[70]} \textsc{UpdateActiveGoals}: goal $G1$ still valid }

    \State{\textbf{[70]} \textsc{SelectIntentions}: plan $P2$ still valid, intention $I2$ still active }

    \State{\textbf{[70]} \textsc{RT-ProgressAndMonitorIntentions}: $I2$($P2$: $C1$ gat\-her\_re\-sour\-ce \& $C2$ deposit\_resource)}

\end{algorithmic}
\end{minipage}
}
\end{execution*}

\paragraph{Reasoning cycle.}
Here, we discuss a step-by-step execution of the agent reasoning cycle, generated by executing the following scenario.
There are two robots C1 and C2 that collaborate to gather resources R1 and R2 and deliver them to the warehouse W.
The Execution \ref{alg:execution_logs} presents an excerpt of the execution logs, in which log lines have the following structure: \textbf{[timestamp]} \textsc{EventName}: additional information.
%

Events reported in the logs include the beginning of a new reasoning cycle \textsc{ReasoningCycle} and its phases, referring directly to the function defined in Algorithm \ref{alg:reasoning_cycle} presented in Section \ref{sec:real-time_bdi_architecture}, which includes goal deliberation \textsc{UpdateActiveGoals}, plan selection \textsc{SelectIntentions}, intention progress and monitoring \textsc{RT-ProgressAndMonitorIntentions}, and sensing \textsc{ReadSensingData}.  Additionally, \textsc{Player’s interaction} identify an interaction of the player with the game.

The first reasoning cycle starts at the beginning of the simulation and ends when the first step in the current intention is completed, in 10 time units, after having sensed the expected effects.
Then, in the middle of the second reasoning cycle, the player spawns a new robot in the scene (time: 15).
When the current intention sub-task ends, at 20 time units, the agent became aware of the new robot.
In the third reasoning cycle, the context condition of the old plan $P1$ does not hold anymore because the total number of available agents changed, so the agent replan its intentions considering now both the robots, then starts to execute parallel sub-tasks, one for each robot.
Finally, after additional omitted steps, sixth reasoning cycle begins and robots collect and deposit resources.

With this example, we provide in-depth details on the agent reasoning cycle, specifically, about their awareness of the surrounding environment and the effects of their own actions, reactivity to external events, and re-planning capabilities.

\robol{
When, in the execution phase, external events prevent the plan to complete its execution, the agent may need to trigger re-planning and \roveri{analyze}{later re-simulate} the schedulability of \roveri{newly generated}{} tasks.
We adopt such a re-planning approach also in the run-time synchronized video-game simulation, so that in the case of an external event or missed deadline a new plan is computed trying to achieve the goal on time.}{}

\paragraph{Deliberation goal deadlines and real-time scheduling constraints.}
Here we compare goal and plan deadlines, scheduling constraints at the deliberation (planning) level, and constraints at the real-time scheduling level.  We consider a scenario with two robots that act (move or collect resources) at the same time, given a plan that include parallel actions.

\begin{figure}[!tb]
    \centering
    \includegraphics[width=0.95\columnwidth]{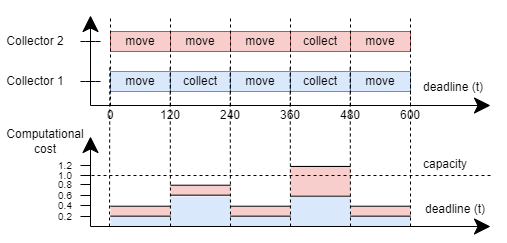}
    \caption{Timeline of the simulation: sequence of parallel
      durative actions for the collector robots (upper part);
      cumulative computation cost of executing such actions in
      parallel (lower part).}
    \label{fig:planning_and_scheduling_timeline}
\end{figure}

\robol{
Figure 2 shows the timeline of a simulation, in which a single-agent system is planning tasks for two controlled robots.
\roveri{}{In the deliberation phase, the agent produces a plan that satisfies deadlines for the achievement of the goal, by assigning parallel actions to the two robots (see  upper part of the figure).
However, in the scheduling phase, when computational cost is taken into consideration, the agent realizes that the cumulative computation cost at step four exceeds the maximum computational capacity available, as shown in the lower part of the figure.}%
}%
\noindent \roveri{ More in detail, }{Figure 2 shows the timeline of the simulation. In }%
the upper part \roveri{of Figure \ref{fig:planning_and_scheduling_timeline} depicts}{is depicted} the sequence of parallel durative actions for the collector robots, as generated by the planner. \roveri{The lower part depicts}{In the lower part, is depicted} the cumulative computation cost of executing such actions in parallel.
In the deliberation phase the planner generates a plan considering action deadlines and assigning at most one single action to each agent at the same time.
The result is a valid optimal plan to achieve the overall goal in just 600 time units.
At a second stage, in the \roveri{execution}{executing} and monitoring layer, the real-time system has to schedule the \roveri{the plan taking into consideration the}{} computational capacity so that each action completes within its own deadline while computational cost does not exceed current computational capacity.
Even if, at the planning stage, no constraints were violated, now, at the real-time scheduling stage, it is not possible to fulfill the intention given the maximum available computational capacity. \roveri{Thus, the agent realizes that the cumulative computation cost at step four exceeds the maximum computational capacity available, and the plan cannot be skeduled.}{}
This is due to the fact that, in a multi-agent environment, some preconditions may be broken by the other agents, therefore the agents may fail to complete their original intention, and this may trigger additional operations to handle this contingency (e.g. generation of new desires or new intentions).

\robol{This example demonstrates that if the planning does not properly consider scheduling aspects it can lead to solutions that could not be executed in a real-time execution environment.}
{This example demonstrates that a misalignment in the configuration of the simulation can lead to inapplicable solutions.}
\robol{More specifically, if the planning assumption is that robots can be assigned each with one action, in parallel with others, this may not be true for the low-level real-time scheduler, for which actions may have varying costs that, when summed, may exceed the maximum available computation capacity.}
{More specifically, we have seen that if the planning assumption is that robots can be assigned each with one action, in parallel with others, this may not be true for the scheduler, for which actions may have varying costs that, when summed, may exceed the maximum available capacity.}

\robol{We are working to overcome this problem by studying a more tight integration of planning and execution where details of the low-level real-time scheduling are taken into account, at planning level, within the search for a solution.}
{As future work, we plan to provide a specific support for this limit case, including guidelines for the users and mapping rules to consider scheduling aspects at the planning stage, so that the planner always generate plans fully schedulable.}


\section{\uppercase{Related Work}}
\label{sec:related_works}
The architecture presented in this article combines multiple technologies and propose a solution that takes advantage of the flexibility of the BDI model in real-time scenarios.


A three-layer autonomy architecture has been first discussed
in~\cite{DBLP:conf/aaai/Gat92,Gat97,Ghal01} and constituted the basis
for several subsequent
works~\cite{Livingstone,Titan,MUROKO,MMOPS,GOACASTRA2011,DBLP:journals/tist/BozzanoCR21}.
Despite these works addressed the problem of providing autonomy to
(single-)agents (mostly in the aerospace domain), their BDI reasoning
capabilities are rather limited, and they lack of a proper handling of
time and real-time constraints. In this work we showed how to decline
it to handle a complete BDI reasoning framework with (real-)time
constraints.

Several frameworks have been proposed in the literature to associate
temporal schedules to BDI agent's tasks (e.g.,
Jade~\cite{bellifemine1999jade}, Jack~\cite{busetta1999jack},
Jason~\cite{bordini2007programming}).
However, none of them explicitly provides an explicit representation
and handling of time and of real-time constraints in the
decision-making and decision actuation processes. The consequence is
that agents overlook the temporal implications of their intended
means, resulting in possible delays in the execution of their tasks,
when an overload of the computation resources occurs.

There are also works that deal with real-time scheduling in a BDI setting (e.g.\cite{DBLP:conf/paams/AlzettaGMC20}, \cite{DBLP:conf/atal/VikhorevAL11}, \cite{DBLP:conf/agents/VincentHLW01} and \cite{DBLP:journals/aamas/CalvaresiCMDNS21}).
In~\cite{DBLP:conf/paams/AlzettaGMC20}, the authors proposed a RT-BDI architecture that, similarly to our one, integrates real-time concepts into the reasoning cycle of a BDI agent.
However, while they can guarantee the respect of deadlines at the task level, the agent's goals agents are not subject to a deadline, and thus they do not allow for the agent taking decisions based on goal's related deadlines. 
Our solution enforces deadlines at all levels, thus ensuring real-time compliance of the entire execution process.
AgentSpeak(RT)~\cite{DBLP:conf/atal/VikhorevAL11} provides logical operators and scheduling criteria, delivered through TAEMS task structures (inherited from AgentSpeak(XL)~\cite{DBLP:conf/atal/BordiniBJBVL02}), associated with deadlines. This framework allows programmers to specify an upper bound for the reaction to events with prioritization criteria.
This approach, similarly to our, deals with deadlines, but differently from us, it does not address the schedulability problem of the selected actions on a real-time operating system and it does not leverage on appropriate computational capacity scheduling algorithms~\cite{PAAM12}. 
%
%
Soft Real-Time~\cite{DBLP:conf/agents/VincentHLW01} focuses on the granularity of deadlines in applications where fractions of seconds are relevant by proposing a specialized scheduler for processes that can deal with hard-deadlines, but still above the grain-size afforded by the operating system, where local competing processes exists.
Differently, we show the deployment on a native real-time operating systems, in which full control over the real-time scheduling of processes is possible to stick to required reaction times and computational costs.
Compared to \cite{DBLP:journals/aamas/CalvaresiCMDNS21}, that provides mostly a formal framework for RT-MAS, the contribution of our paper consists in going down to the scheduling of tasks/actions of a plan on
the processor. 
We further remark that, w.r.t. all the above mentioned works, in our framework, we not only considers goal deadlines, but we take into account issues related to dispatchability of the actions of the plans considering the respective execution in a suitable real-time operating environment (e.g. a real-time operating system using a single processor).
Moreover, we also support autonomous deliberation leveraging on the integration of a temporal planner. This way the agent can not only execute pre-defined plans but has the capability of generating new plans on his own from the current situation and his current beliefs to achieve desires.
Finally, our architecture considers both scheduling constraints and goal deadlines, thus supporting full real-time control over the whole execution\roveri{, and as far as we know, these}{ (as far as our knowledge is concerned, this} are unique features of our architecture\roveri{}{)}.


In the video-game setting, several works adopted BDI architectures mainly relying on machine learning techniques, and few ones use model-based approaches.
\cite{Bartheye2009,Bartheye_Jacopin_2009} show how PDDL planning can be used to implement the concept of a BDI agent inside the \emph{Iceblox} and in the \emph{VBS2} games.
%
%
All these solutions simply let the BDI agent complete the levels by
its own, avoiding the main challenges: the unpredictability of the
player's action and the handling of real-time constraints.


As far as autonomous deliberation is concerned, we remark that typically agent's programming languages tend to not support this feature, leading to the necessity of creating ad-hoc systems to implement it. 
\cite{meneguzzi_de_silva_2015} claims the general preference in implementing BDI agents through the usage of predefined plan libraries to reduce the computational cost, rather than integrating automatic deliberation reasoning capabilities.
In our architecture, we leverage a PDDL temporal planner to generate new plans to deal with contingencies when flexibility is fundamental. The new plans can be added to a plan library for future re-use.
%


\section{\uppercase{Conclusion}}
\label{sec:conclusion}
\roveri{In this paper we addressed the problem of developing agents
that can reason and act taking into account the fact that the agents
reason and execute on top of a real-time infrastructure with limited
computation capabilities.
The proposed RT-BDI-based architecture ensures predictability of
execution, considers as primitive citizens concepts like computational
capacity, deadline, scheduling constraints, durative actions, periodic
tasks and temporal planning deliberation.
The proposed architecture is structured in three layers: the BDI
layer, responsible for handling belief, desires (goals) and intention
(temporal plans); the Execution and Monitoring layer, responsible for
executing and monitoring plans; and, the Real-Time layer that handles
the low-level execution of tasks following the scheduling policy
established in above layers.
The proposed framework allows an agent to be aware of time constraints
while deliberating and not only during the execution.
We demonstrated the practical applicability of the proposed framework
through an implementation built on top of the Kronosim BDI simulator,
and we integrated i) the framework with a multi-platform game engine
based on Unity, ii) the OPTIC temporal planner for the deliberation.
}{}

As future work, we plan to deploy our architecture in different
contexts e.g. in the robotic setting by systems extending state-of-art
functionalities with abilities to reason about goals and plans in
real-time. \roveri{We also plan to investigate a more tight
integration of planning and execution where details of the low-level
real-time scheduling are taken into account, at planning level, within
the search for a solution, by e.g., considering to extend OPTIC to
support PDDL 3.1~\cite{pddl31} features.}


\bibliographystyle{apalike}
\begin{small}
\bibliography{main}
\end{small}


\appendix
\section{Real-Time BDI Agents: a model and its implementation: Appendices}

This appendix has been provided for the review process and i will be removed upon acceptance and added as external link. This appendix is tructured as follows. Appendix \ref{sec:a1} presents additional details of the Kronity video-game based on Unity and powered by our simulator. Appendix \ref{sec:a2} presents complemented with a graphical representation the details of the architecture of the developed RT-BDI software simulator. Finally, Appendix \ref{sec:a3} presents the additional scenarios we considered for the validation of the proposed architecture.

\subsection{Use case scenario: Kronity}
\label{sec:a1}

As application scenario for our framework, we implemented a resource collection video game called \textbf{Kronity}, where
\begin{itemize}
    \item the concept of non-player character (NPC) has the same purpose as the BDI agent model, which is to emulate, to some degree, the human behavior and reasoning process;
    \item the player interaction with the game provides an optimum simulation of an unpredictable environment;
    \item the video game editors (Unity in our case) can provide the tools to implement the visualization aspect of our system.
\end{itemize}


\begin{figure}[!htb]
    \centering
    \includegraphics[width=0.95\columnwidth]{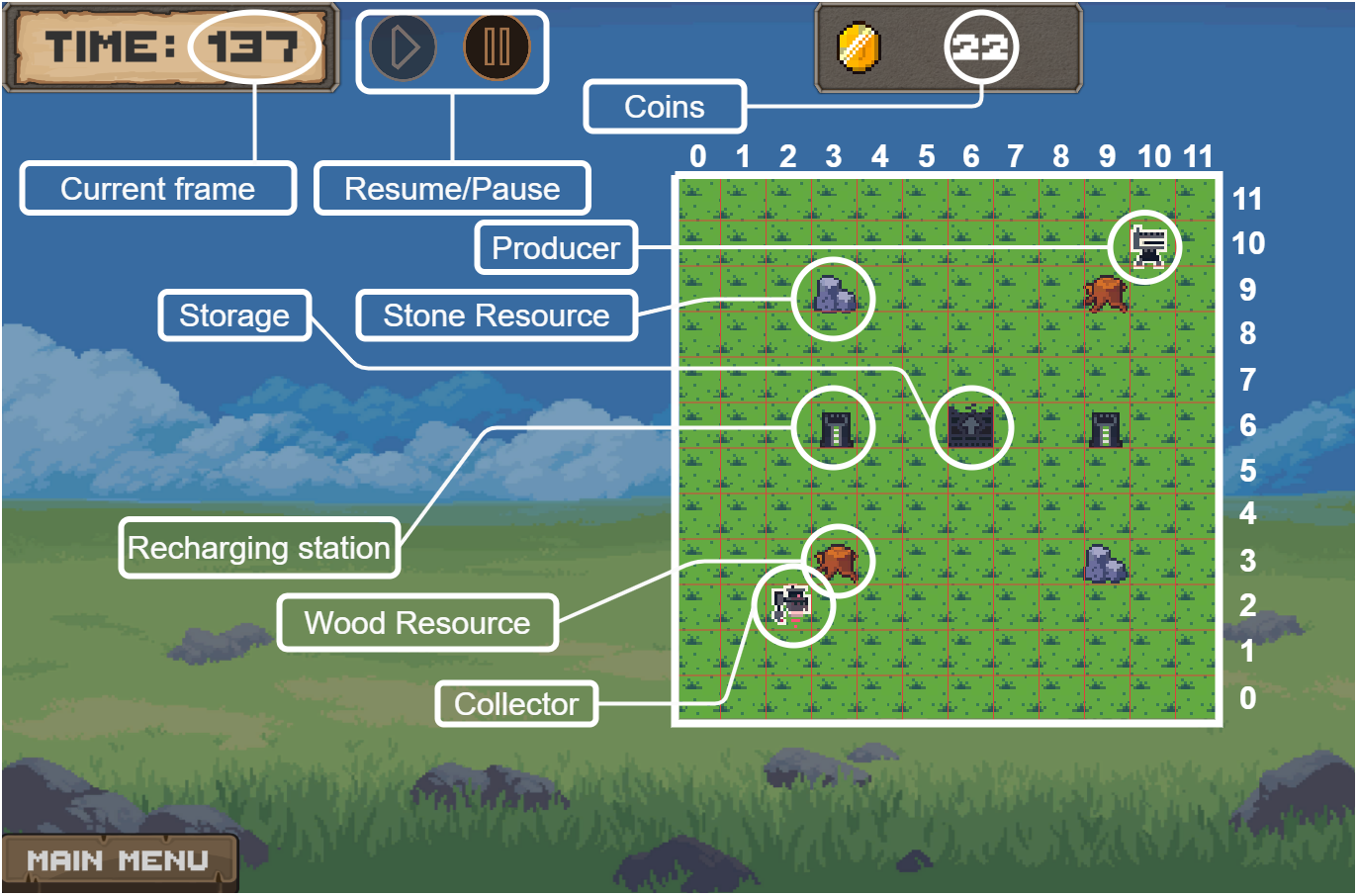}
    \caption{Screenshot taken from the main section of Kronity.}
    \label{fig:game_screenshot}
\end{figure}

Kronity can be defined as both a puzzle and a real-time strategy game, and it is composed by two parts: the problem generator and the main section.
In the problem generator, the player can select one of the proposed predefined levels. A level defines part of an initial state, that must be completed by the player based on the solution it thinks is the best, and a final state. An example of an initial state is represented in Figure \ref{fig:game_screenshot}.
In the main section, instead, the game executes the sequence of actions needed to transition from the initial state to the final one. The player can interfere with the execution by moving the robots or changing some of their values: the game will then autonomously try to find a new solution to achieve the goal. However, these player's actions have a cost in terms of in-game currency, defined as "coins" in Figure \ref{fig:game_screenshot}. The map is built as a square grid, logically divided into cells that can be both empty or occupied by any number of entities.

While all entities have knowledge of the cell they are positioned into, many other information could be different based on their role inside the game. These roles are the following:
\begin{description}
    \item[Resources:] static entities that indicate where a certain resource can be found in the environment.
    \item[Robots:] the only entities which are capable of moving around in the environment; hence, they are the primary actors of the game, those which make it progress. All robots have information about the number of resource samples they carry and about the amount of charge of their own battery. There are two different kinds of robot implemented in the game:
      \begin{itemize}
      \item \textbf{Collectors}, which are exclusively able to collect resources distributed in the environment.
      \item \textbf{Producers}, which are exclusively able to combine raw resources into new ones (only producers can hold these combined resources).
      \end{itemize}
    Furthermore, robots can exchange raw resources with each other.
    \item[Storage:] static entities that indicate where resources can be stored and hold the information about how many resources are currently stored; usually, the storage is the landmark in the creation process of the goal (e.g., as a final state $x$ samples of the wood-type resource must be stored inside the storage).
    \item[Recharging Stations:] static entities that let robots recharge their battery in order to continue with their job.
\end{description}

\subsection{Software architecture of the real-time BDI agents simulator}
\label{sec:a2}
In this section we provide a graphical representation of the simulator software architecture presented in Section 4 and based of an extension of Kronosim.

\begin{figure}[!h]
    \centering
    \includegraphics[width=\columnwidth]{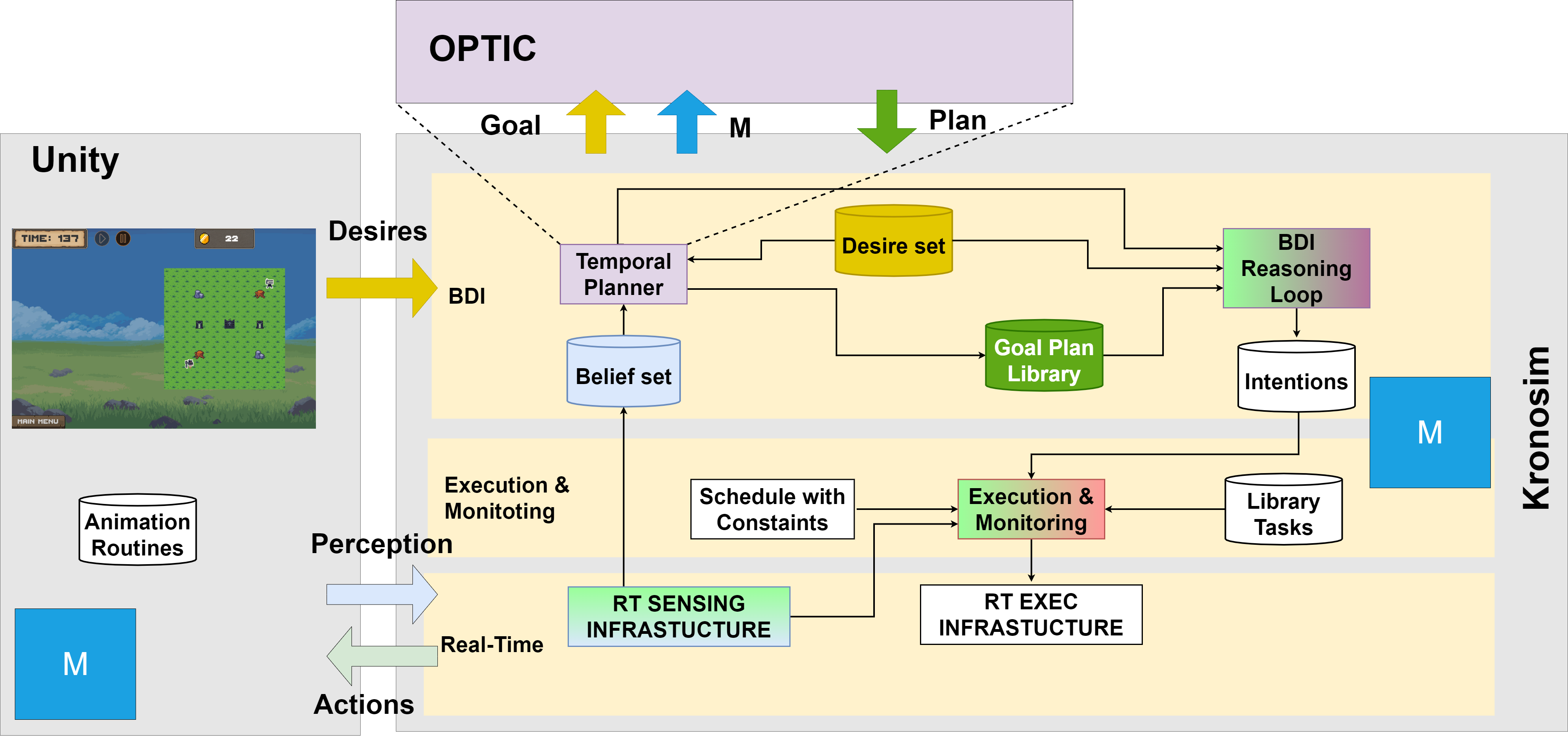}
    \caption{Software architecture of the Real-Time BDI Agents simulator based on Unity, Kronosim and OPTIC.}
    \label{fig:architecture}
\end{figure}

Figure~\ref{fig:architecture} provides a pictorial representation of the deployment of the proposed three layers RT-BDI architecture, with the three main software components: Kronosim, Unity and OPTIC.

On the left of the figure, Unity, adopted as game engine, provides the simulation of the game including a graphical user interface and graphical animations. In addition, it provides APIs to communicate with the BDI agents, i) by forwarding high-level goals given by the user to agents, in the form of desires, ii) notifying changes in the enviroment to agents through perceptions, and iii) receiving low-level commands from the agents in the form of actions (from PDDL action).
The blue box in the bottom left part, represents the mapping components used to map concepts between Kronosim and Unity.

In the center/right side of the figure, Kronosim, adopted to simulate a real-time agent, provides the BDI reasoning cycle extended with planning and scheduling capabilities, structured in threee layers, as detailed in Section 3.

Finally, in the upper part of the figure, OPTIC, provides planning capabilities to Kronosim, by receiving a goal together with a model of the problem and sending back a computed solution plan.

\subsection{Validating the RT-BDI Agent Architecture: additional scenarios}
\label{sec:a3}
In this section we integrate the scenarios discussed in the paper in Section 5 with three additional ones: i) reactivity to external events, ii) coordinator agent, and iii) learning agents.

\subsubsection{Reactivity to external events}

We present here a scenario that considers the interaction of the
player with the video-game, perceived as an external event by the
agent who reacts by re-planning its current intention.  The scenario
discussed here considers a collector robot who tries to reach the
recharging station to refill its battery, while the player moves it on
the other side of the map.

\begin{figure}[!htb]
    \centering
    \includegraphics[width=0.95\columnwidth]{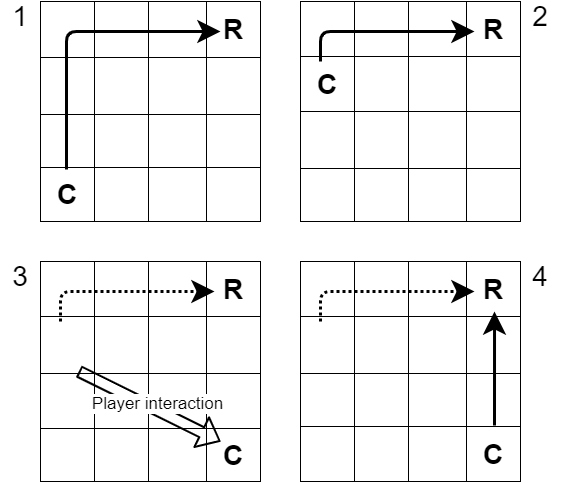}
    \caption{Schematized representation of a simple execution of the game. The figure is divided into four sectors that represents as many different moments of the execution itself.}
    \label{fig:game_schema}
\end{figure}

Figure \ref{fig:game_schema} presents the scenario with a schema of
the game in four different time instants of execution.  In the first
time instant (top left part of the figure), the collector "C", who
wants to reach the recharging station "R" to refill its battery, has a
navigation plan represented by the arrow.
Later, in the second time instant (top right part of the figure), the
collector robot "C" is half the way its path to the recharging
station.
Then, in the third time instant (bottom left part of the figure), the
player interferes with the execution, moving the collector robot out
of its path. The agent perceives and deliberates on this only when the
current movement action (single cell movement) is completed.
Finally, in the fourth time instant (bottom right part of the figure),
the agent, after perceiving the new unexpected position, submits a new
intention to navigate towards the charging station from the new
position. If a plan is not readily available in the agent's knowledge
base (Goal Plan Library), the generation of a new one is triggered in
the planner.

This scenario demonstrates a simple example of reactivity of the
agent, to support external events, such as, human interaction.
In the current implementation, agent reactivity is limited by the
execution of scheduled tasks, which needs to execute completely before
the agent can deliberate on the new changed situation.

\subsubsection{Coordinator agent}

This scenario discusses about the coordination of multiple agents in a
unified environment.
In our Unity-based video game built on top of our architecture,
scenarios involving multiple agents are simulated adopting a
coordinator agent that dispatches goals directly to the agents, to
orchestrate them, without them having any possibility to communicate
directly one to another.

\begin{figure}[!htb]
    \centering
    \includegraphics[width=0.95\columnwidth]{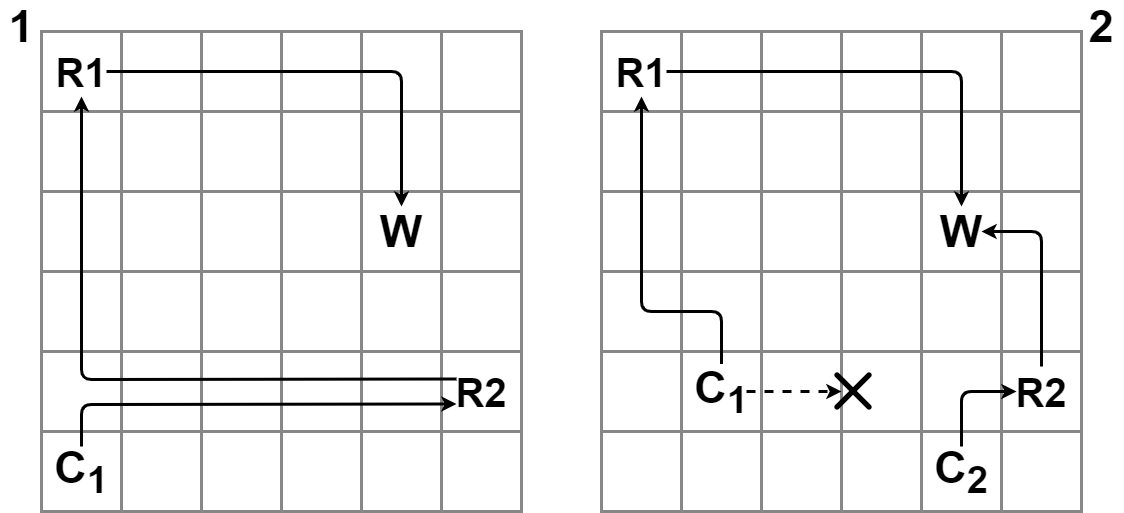}
    \caption{Coordination scenario. Robots C1 and C2 are coordinated by the agent to gather resources R1 and R2 and deliver them to the warehouse W. This scenario shows how the system is able to support multiple agents and split the work-load to optimize the achievement of goals.}
    \label{fig:MAS}
\end{figure}

Figure \ref{fig:MAS} shows a practical example of coordination.
In this example, the overall goal of the game, pursued by the
coordinator agent, is to deliver all the resources to the warehouse.
Initially, the robot C1 is assigned to gather both resources R1 and R2
and deliver them to warehouse W.
To do this the robot has a known plan available that is showed on the
left of Figure \ref{fig:MAS}.
Then, a second robot is added to the scene.  The coordinator agent
perceives this information and triggers a new execution of the control
loop: (i) the belief set is updated, (ii) the goal of delivering all
the resources to the warehouse is evaluated, (iii) a new plan that
consider both robots is computed to optimize the outcome, (iv) the
robots are assigned with new goals and associated specific plans that
satisfy the goal without having the robots interfering one to another.
On the right of Figure \ref{fig:MAS} are shown the plans used by the
two robots to achieve the goal.  As in the case of the scenario
represented in the figure, it can be the case that robots are
reassigned with different goals and need therefore to interrupt the
intentions in execution.  In this case, robot C1 is rerouted towards
resource R1, ignoring its previous task R2, while the newcomer robot
C2 is assigned to gather resource R2.

The final outcome of the coordination is that the overall goal
(i.e. having resources delivered to the warehouse) is achieved by
employing both robots to speed up the process but still avoiding them
to interfere one to another.
This scenario demonstrates how it may be possible to scale up our
architecture by aggregating agents into coordinated units.
In complex scenarios, single agents may be de-composed into units of
agents that act in a coordinated way in order to achieve the agent
main goal.

\subsubsection{Learning agents }

In this section we discuss about the ability of agents to learn new
plans and reuse them later on by adding them to the knowledge-base
(Goal Plan Library).  We consider a scenario in which an agent misses
a plan for a sub-goal and uses the planner to generate it, and achieve
the overall goal.  This behaviour is supported by different features
of the agent: planning capabilities, plan de-composition into
sub-goals, and growing of the plans knowledge base.

\begin{figure}[!htb]
    \centering
    \includegraphics[width=1.00\columnwidth]{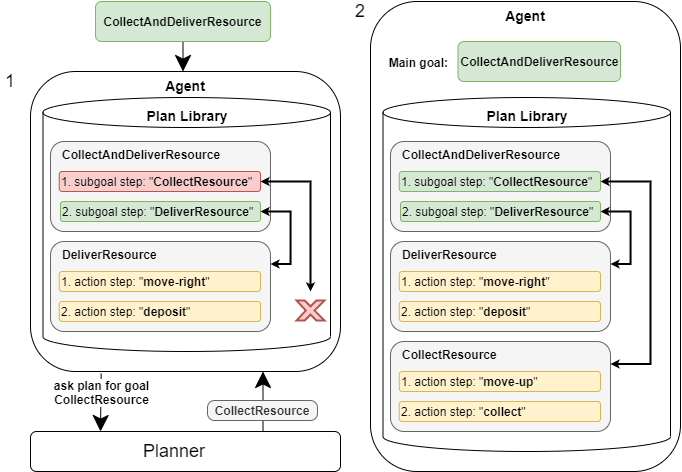}
    \caption{Evolution of the agent internal status, including plans knowledge base and current intention.}
    \label{fig:agent_internal_state}
\end{figure}

Figure \ref{fig:agent_internal_state} describes the evolution of the
agent internal status, including plans knowledge base and current
intention.
In a first time instant (left part of the figure) the agent, asked to
pursue the goal "CollectAndDeliverResource", has a plan that consists
of two sub-goals, of which the first completely misses an applicable
plan.  The planner is therefore invoked to get a plan for the first
sub-goal, which can then be added to its knowledge base.
Then, in a second time instant (on the right part of the figure) the
agent has learned all necessary plans for goals and sub-goals, and can
start to fulfill intentions to achieve "CollectAndDeliverResource".

This scenario demonstrates how agents can learn and store new plans
and re-use them later in the future.  Note that, although our
simulator features (i) planning capabilities, (ii) plan de-composition
into sub-goals, and (iii) growing plans knowledge base; in the current
implementation of the simulator, planning capabilities are only
supported at main goal level, and the planner cannot be called to
generate partial sub-goal plans. This limitation is only for the sake
of simplicity of the prototype implementation. Indeed, nothing
prevents invoking the planner to generate partial sub-goal plans (this
is left for future work).




\end{document}